\documentclass[prb,amssymb,twocolumn]{revtex4}
\usepackage{graphicx}

\begin{document}

\title{Quantum Spin Hall Effect in Silicene}

\author{Cheng-Cheng Liu}
\affiliation {Beijing National Laboratory for Condensed Matter Physics and Institute of Physics, Chinese Academy of Sciences,Beijing 100190, China}

\author{Wanxiang Feng}
\affiliation {Beijing National Laboratory for Condensed Matter Physics and Institute of Physics, Chinese Academy of Sciences,Beijing 100190, China}

\author{Yugui Yao$^*$}
\affiliation {Beijing National Laboratory for Condensed Matter Physics and Institute of Physics, Chinese Academy of Sciences,Beijing 100190, China}

\maketitle

{\bf Recent years have witnessed great interest\cite{kane2005a,PhysRevLett.95.146802,Science.314.1757,Science.318.766,PhysRevLett.97.236805,PhysRevLett.100.236601} in the quantum spin Hall effect (QSHE) which is a new quantum state of matter with nontrivial topological property due to the scientific importance as a novel quantum state and the technological applications in spintronics. Taking account of Si, Ge significant importance as semiconductor material and intense interest in the realization of QSHE for spintronics, here we investigate the spin-orbit opened energy gap and the band topology in recently synthesized silicene using first-principles calculations. We demonstrate that silicene with topologically nontrivial electronic structures can realize QSHE by exploiting adiabatic continuity and direct calculation of the $Z_2$ topological invariant. We predict that QSHE in silicene can be observed in an experimentally accessible low temperature regime with the spin-orbit band gap of $1.55$ meV, much higher than that of graphene  due to large spin-orbit coupling and the low-buckled structure. Furthermore, we find that the gap will increase to $2.90$ meV under certain pressure strain. Finally, we also study germanium with similar low buckled stable structure, and predict that SOC opens a band gap of $23.9$ meV, much higher than the liquid nitrogen temperature.}

Quantum spin Hall effect (QSHE) with time reversal invariance is gapped in the bulk and conducts charge and spin in gapless edge states without dissipation at the sample boundaries. The existence of QSHE was first proposed in a graphene in which the spin-orbit coupling (SOC) opens a band gap at the Dirac point by Kane and Mele\cite{kane2005a}. But the subsequent work found the SOC rather weak, which is in fact a second order process for a flat graphene, so the QSHE in a flat graphene can only occur at unrealistically low temperature\cite{yao2007,PhysRevB.74.155426,PhysRevB.74.165310}. So far, there is only one proposal that is able to demonstrate QSHE in a real system, which is in two dimensional mercury telluride-cadmium telluride semiconductor quantum wells\cite{Science.314.1757,Science.318.766} in despite of some theoretic suggestions\cite{PhysRevLett.97.236805,PhysRevLett.100.236601}. Nevertheless, both graphene and HgTe quantum wells are not good enough to be compatible with the present silicon-based electronics industry. As the counterpart of graphene\cite{nature.materials.VOL.6.MARCH.2007} for silicon, silicene has been shown that a low buckled two-dimensional hexagonal structure corresponds to a stable structure, and there are also evidences of graphene-like electronic signature in silicene nanoribbons experimentally\cite{Aufray2010,Padova2010,Appl.Phys.Lett.97.223109}. Therefore, almost every striking exceptional property of graphene could be transferred to this innovative material with the extra advantage of easily being incorporated into the silicon-based microelectronics industry.

\begin{figure}
\includegraphics[width=3.5in]{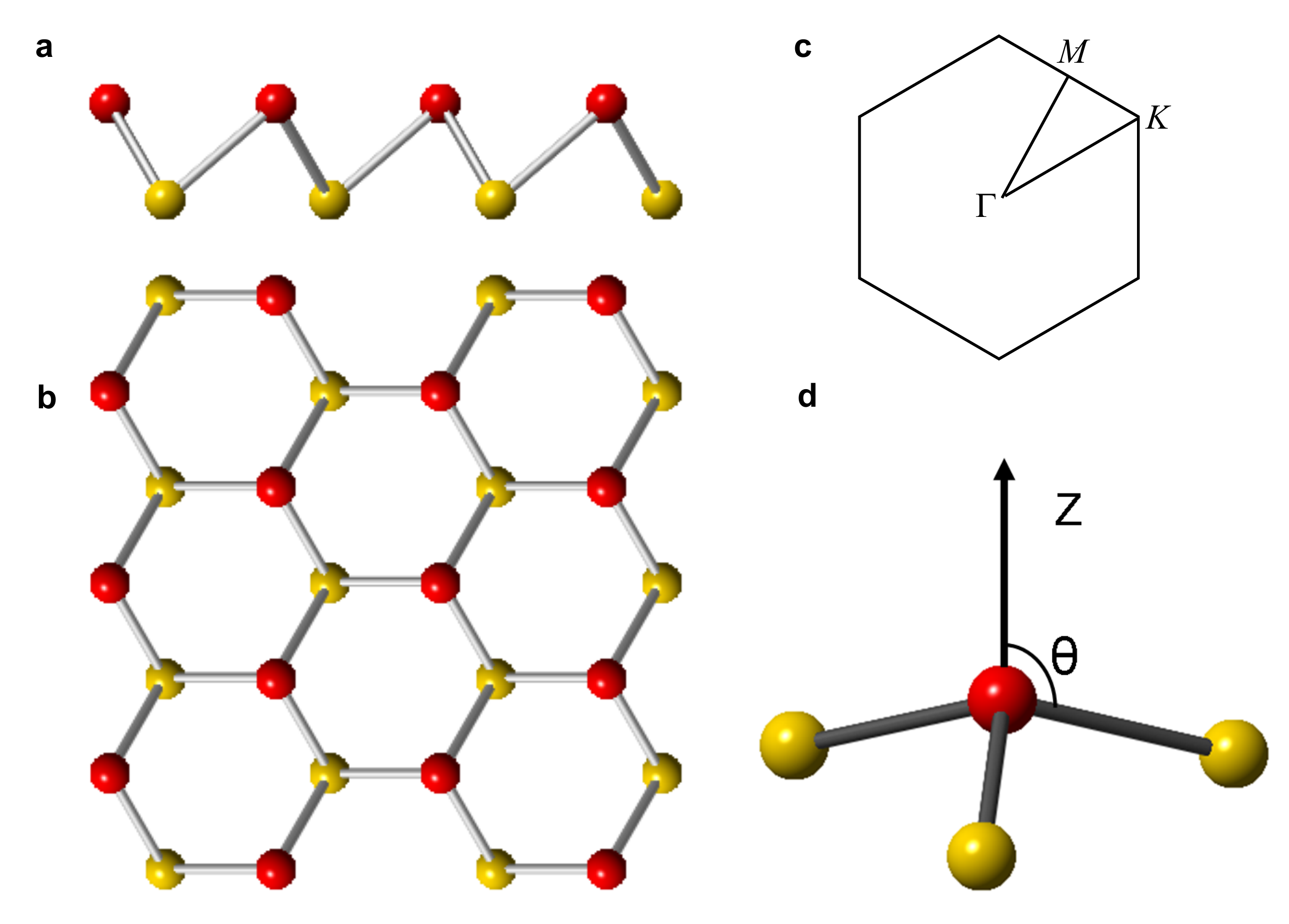}
\caption{\textbf{The lattice geometry of low buckled silicene.} \textbf{a ,b,} The lattice geometry from side view and top view respectively. Note that A-sublattice (red) and B-sublattice (yellow) are not coplanar. \textbf{c,} The first Brillouin zone of silicene and its points of high symmetry. \textbf{d,} The angle $\theta$ is defined as between the Si-Si bond and the $Z$ direction normal to the plane.}\label{fig:geometry}
\end{figure}

The structure of silicene is shown in Fig.~\ref{fig:geometry}. We obtain the low-buckled geometry of minimum energy and stability with lattice constant $a=3.86$ \AA \ and nearest neighbor Si-Si distance $d=2.28$ \AA \ through structural optimization and calculations of phonon spectrum. The results agree with the previous work\cite{Cahangirov2009}. Compared with graphene, the larger Si-Si interatomic distance weakens the $\pi-\pi$ overlaps, so it cannot maintain the planar structure anymore. This results in a low-buckled structure with $sp^{3}$-like hybrid orbitals. In Fig.~\ref{fig:geometry}, one can define the angle $\theta$ between the Si-Si bond and the direction normal to the plane. The $sp^{2}$ (planar), low-buckled and $sp^{3}$ configurations correspond to $\theta=90$, $\theta=101.73$ and $\theta=109.47$ degrees respectively.

To illustrate the band topology of the low-buckled silicene, we begin with their graphene analog, planar silicene, and follow its band structure under an adiabatic transformation during which the unstable planar honeycomb structure is gradually evolved into the low-buckled honeycomb structure. Planar silicene with the same structure as graphene should have the similar properties.  Furthermore, since Si atoms have greater intrinsic spin-orbit coupling strength than C atoms, it is natural to conceive that the quantum spin Hall effect is more significant in planar silicene. According to symmetry, the low energy effective Hamiltonian with SOC in silicene in the vicinity of Dirac point $K$ can be described by
\begin{eqnarray}
&&H^{[K]}_{eff}\approx\left(
\begin{array}{cc}
\xi\sigma_{z}&v_{F}(k_{x}+ik_{y})\\
v_{F}(k_{x}-ik_{y})&-\xi\sigma_{z}\
\end{array}
\right)\label{soeff}
\end{eqnarray}
where $v_{F}$ is the Fermi velocity of $\pi$ electrons near the Dirac points with the almost linear energy dispersion,and $\sigma_{z}$ is Pauli matrix. The effective SOC $\xi$ for planar silicene has the explicit form $\xi\approx2\xi_{0}^{2}|\Delta_{\epsilon}|/(9V_{sp\sigma}^2)$ with $\Delta_{\epsilon}$ being the energy difference between the 3s and 3p orbitals and $\xi_{0}$ the intrinsic spin-orbit coupling strength respectively. The parameter $V_{sp\sigma}$ corresponds to the $\sigma$ bond formed by the 3s and 3p orbits. The effective Hamiltonian near Dirac point $K^*$ can be obtained by the time-reversal operation on the one near $K$. The above equation results in a spectrum $E(\vec{k})=\pm\sqrt{(v_{F}k)^{2}+\xi^{2}}$. Therefore, one can estimate the energy gap, which is $2\xi$ at the Dirac points, to be about the order of $0.1$ meV by taking the values of the corresponding parameters\cite{Harrison}. Notice that in planar silicene ($\theta=90$) $\pi$ orbitals and $\sigma$ orbitals are coupled only through the intrinsic SOC. So, the effective SOC is in fact a second order process. However, with the deviation ($\theta>90$) from the planar geometry, $\pi$ orbitals and $\sigma$ orbitals can also directly hybridize. Consequently, the magnitude of the effective SOC depends on the angle $\theta$. As can be expected with increasing the degree of deviation from the planar structure, the effective SOC will be incremental, and QSHE will be more significant.

The argument above is supported by our first-principles calculations based on density-functional theory(DFT). The relativistic electronic structure of silicene is obtained self-consistently by using the projector augmented wave (PAW) pseudopotential method implemented in the VASP package\cite{PhysRevB.54.11169}. The exchange-correlation potential is treated by Perdew-Burke-Ernzerhof (PBE) potential\cite{PhysRevLett.77.3865}.

\begin{figure}
\includegraphics[width=3.5in]{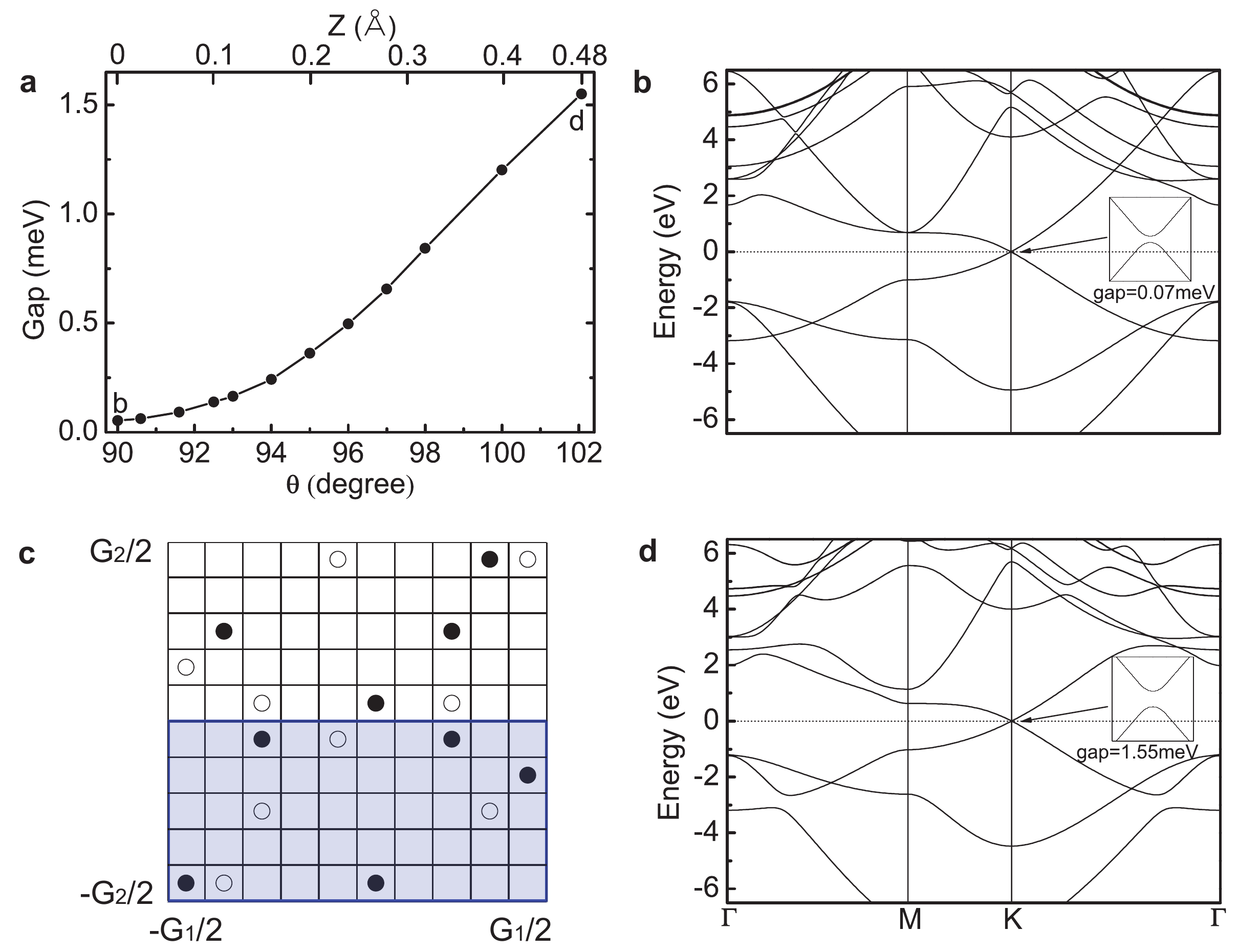}
\caption{\textbf{The adiabatic evolvement of the gap, calculated relativistic band structure and the topological $Z_2$ invariant of silicene.} \textbf{a,} The evolvement of the gap opened by SOC for the $\pi$ orbital at the Dirac point $K$ from the planar honeycomb geometry to the low-buckled honeycomb geometry with keeping Si-Si bond length constant. The top and down abscissas correspond to the difference of $Z$ abscissa between A-sublattice and B-sublattice and the $\theta$ angle aforementioned respectively during evolvement. \textbf{b} and \textbf{d} are the relativistic band structures with the corresponding geometries in \textbf{a}. \textbf{b, d,} Main panel: the relativistic band structure of planar silicene and low-buckled  silicene respectively. Inset: zooming in the energy dispersion near the $K$ point and the gap induced by SOC. \textbf{c,} The $n$-field configuration for silicence. The calculated torus in Brillouin zone is spanned by $G_1$ and $G_2$. Note that the two reciprocal lattice vector form a angle of 120 degrees. The white and black circles denote $n=1$ and $-1$, respectively, while the blank denotes $0$. The $Z_2$ invariant is $1$ obtained by summing the $n$-field over half of the torus.}\label{fig:gapevolveandbs}
\end{figure}

We carry out detailed and systematic calculations of the band structure in adiabatic evolvement from the planar honeycomb geometry to the low-buckled honeycomb geometry. The evolvement of the gap opened by SOC for the $\pi$ orbital at the Dirac point K from the planar honeycomb geometry to the low-buckled honeycomb geometry is shown in Fig.~\ref{fig:gapevolveandbs}a. Fig.~\ref{fig:gapevolveandbs}b and Fig.~\ref{fig:gapevolveandbs}d show the band structures of planar and low-buckled silicene respectively with the corresponding structures in Fig.~\ref{fig:gapevolveandbs}a. The band structures of planar and low-buckled geometry are slightly different in consideration of that the gap induced by the effective SOC increases and the degeneracies at some $k$ points split. The difference of the both band structures along $\Gamma$M high symmetric orientation and in the energy range from $-3$ eV to $-2$ eV actually means that $\sigma$ orbital and $\pi$ orbital can directly hybridize only in low-buckled geometry. We can find that the gap induced by SOC for $\sigma$ orbitals is $34.0$ meV at $\Gamma$ point in both geometries. As is also shown in the figure that the magnitude of the gap induced by effective SOC for the $\pi$ orbital at the $K$ point in planar geometry is $0.07$ meV, which is in agreement with the estimate obtained from the tight-binding model discussed above. In low-buckled structure, the magnitude of the gap is $1.55$ meV, which corresponds to $18$ K. The top coordinate in Fig.~\ref{fig:gapevolveandbs}a denotes the distance of two nonequivalent Si atoms within a primitive cell in the vertical direction. The figure indicates that the gap with the magnitude of $0.07$ meV in the planar structure has been increasing to $1.55$ meV in the low-buckled structure with energy minimum and stability. Most importantly, the gap is not closed. Therefore, the low-buckled silicene with energy minimum and stability must share the same nontrivial topological properties as the planar silicene. Consequently, QSHE can be realized in the low-buckled silicene, namely the native geometry of silicene. The argument can be confirmed by direct calculation of the $Z_2$ topological invariant.

One can interpret nonzero topological $Z_2$ invariant as an obstruction to make the wave functions smoothly defined over half of the entire Brillouin zone under a certain gauge with the time reversal constraint\cite{PhysRevB.74.195312,PhysRevB.75.121306,RevModPhys.82.1959}. The band topology can be characterized by the $Z_2$ invariant. $Z_2=1$ characterizes a nontrivial band topology while $Z_2=0$ means a trivial band topology. Here we follow the method in Ref.~\onlinecite{JPSJ-76-053702} to directly perform the lattice computation of the $Z_2$ invariants from our first-principles method\cite{PhysRevLett.105.096404,PhysRevLett.106.016402}. The $n-$field configuration for the low-buckled silicene is shown in Fig.~\ref{fig:gapevolveandbs}c from first-principles calculations. It should be noted that different gauge choices result in different $n$-field configurations, however the sum of the $n$-field over half of the Brillouin zone is gauge invariant module 2, namely $Z_2$ topological  invariant. As shown in Fig.~\ref{fig:gapevolveandbs}c, low-buckled silicene has nontrivial band topology with the topological invariant $Z_2=1$. Therefore, QSHE can be realized in the low-buckled silicene, that is the native geometry of silicene.

\begin{figure}
\includegraphics[width=3.5in]{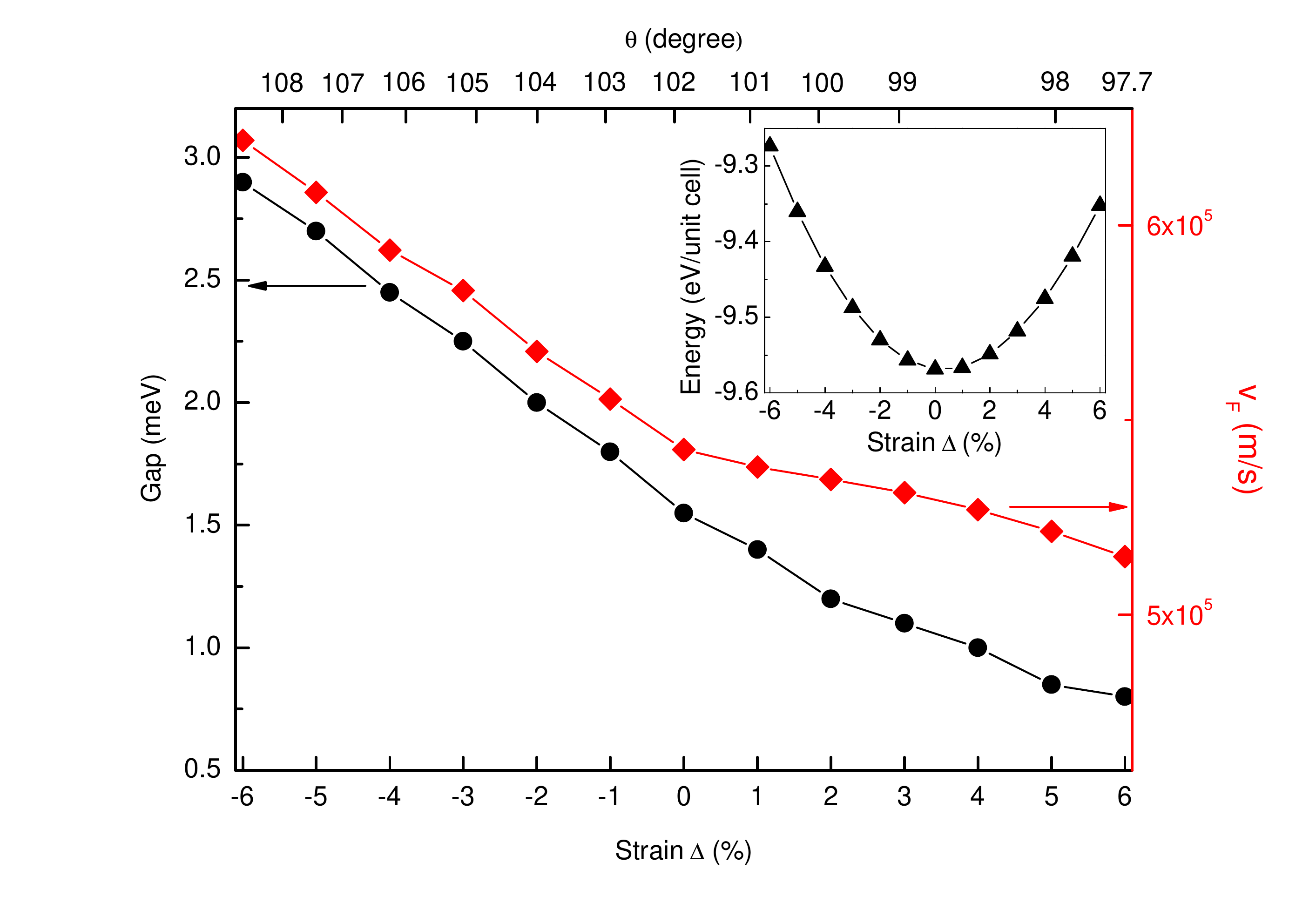}
\caption{\textbf{The gap induced by SOC and the Fermi velocity of charge carriers $v_{F}$ near the Dirac points are calculated from first-principles method under different hydrostatic strain condition.} The black circles and red diamonds mark the gap and the Fermi velocity $v_{F}$ near the Dirac points under different hydrostatic strain $\Delta$. Inset: Energy of unit cell versus different hydrostatic strain condition.}\label{fig:hydrostatic}
\end{figure}

In what follows, we investigate the gap opened by SOC at Dirac points related to QSHE and the Fermi velocity of charge carriers $v_{F}$ near the Dirac points in a series of silicene geometries under hydrostatic strain from first-principles method. We find that while the largest pressure strain can reach $-6\%$ without destroying the nontrivial topological properties of those systems, the magnitude of the gap at Dirac points induced by SOC can be up to $2.90$ meV, which corresponds to $34$ K. As shown in Fig.~\ref{fig:hydrostatic}, the magnitude of the gap at Dirac points induced by SOC is incremental with the decrease of hydrostatic strain $\Delta$, which is defined as $\Delta=(a-a_0)/a_0\times100\%$, where $a_0$ and $a$ being the lattice constant without and with hydrostatic strain respectively. In the pressure strain range, the QSHE can be also realized in the system and even more pronounced. The figure also indicates the greater the angle $\theta$, the greater the gap. In addition, we evaluate the Fermi velocity of charge carriers $v_{F}$ near the Dirac points under different hydrostatic strain and find that the magnitude of the hydrostatic strain does not significantly change the carrier Fermi velocity $v_{F}$. The value is slightly less than the typical value in graphene, say, $10^{6}$ m/s due to the larger Si-Si atomic distance.

Recently, several experiments on silicene have been reported\cite{Aufray2010,Padova2010,Appl.Phys.Lett.97.223109}. They have not only proven silicene adopting slightly buckled honeycomb geometry and possessing the band dispersion with a behavior analogous to the Dirac cones of graphene but also synthesized a silicene sheet through epitaxial growth. With the advancement in experimental techniques, we expect that silicene with high-qualify will soon be manufactured. The experimental data available can be compared with our theoretical prediction then.

\begin{figure}
\includegraphics[width=3.5in]{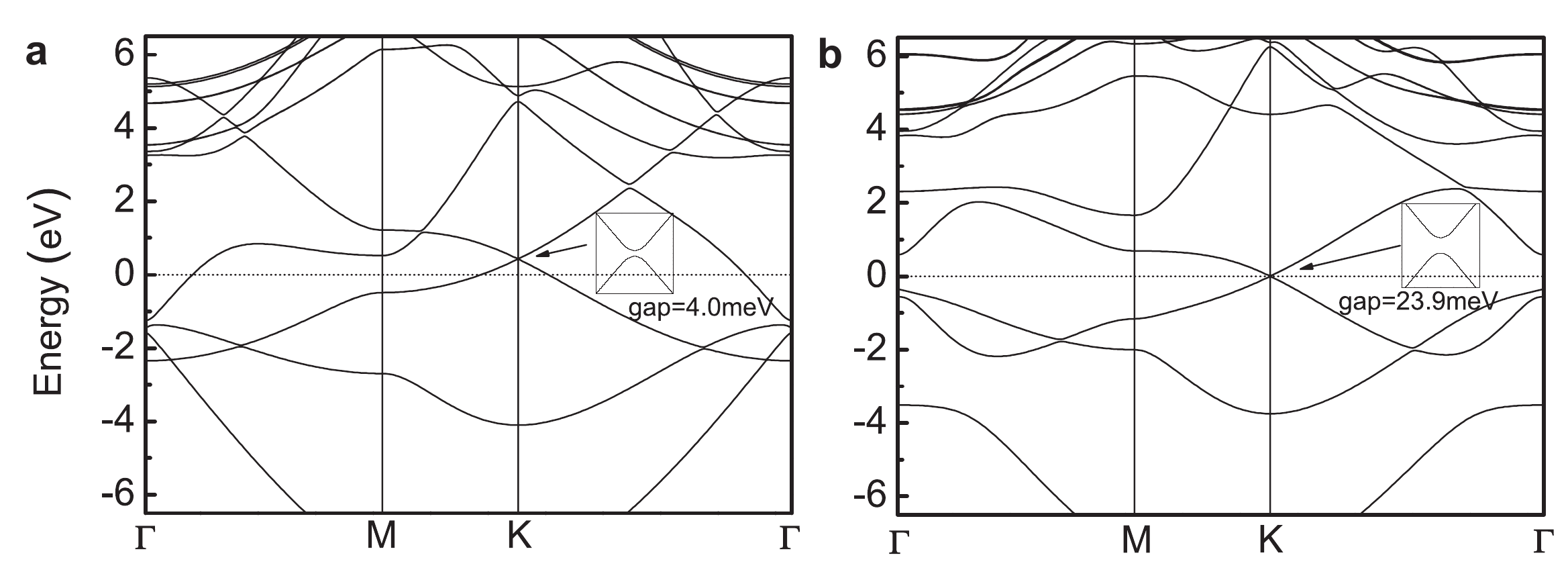}
\caption{\textbf{The calculated relativistic band structure of germanium with honeycomb structure.} \textbf{a, b,} Main panel: the relativistic band structure of Germanium with planar and low buckled honeycomb structure respectively. Inset: zooming in the energy dispersion near the $K$ point and the gap induced by SOC.}\label{fig:GegapZ2}
\end{figure}

Although Germanium with two dimensional honeycomb geometry has not been yet synthesized in experiments so far, we also conduct a detailed study on germanium with two dimensional honeycomb structure because of its similarity to the other group IVA elements in the periodic table, as well as its significant importance as semiconductor material. After structural optimization and calculations of phonon spectrum, the low-buckled geometry of minimum energy and stability with lattice constant $a=4.02$ \AA \ and nearest neighbor Ge-Ge distance $d=2.42$ \AA \ is obtained. As shown in Fig.~\ref{fig:GegapZ2}a and Fig.~\ref{fig:GegapZ2}b, Ge with low-buckled honeycomb structure is insulator while Ge with planar honeycomb structure is metallic. Fig.~\ref{fig:GegapZ2}b indicates that the magnitude of the gap induced by effective SOC for the $\pi$ orbital at the $K$ point in low-buckled geometry is $23.9$ meV corresponding to $277$ K which is much higher than the liquid nitrogen temperature. The direct calculation for topological $Z_2$ invariant proves that Ge with low-buckled honeycomb structure has nontrivial band topology. Therefore, we predict that QSHE will be realized in native germanium with two dimensional low-buckled honeycomb geometry and easily observed experimentally once this novel material is synthesized.

In conclusion, we have shown both silicene and Ge with two dimensional honeycomb geometry have nontrivial topological properties in their native structure. In addition, the QSHE in silicene can be more significant under a range of hydrostatic strain due to the increasing gap size. These are confirmed by direct calculations of the topological $Z_2$ invariants from first-principles methods. Silicene and Ge with the low buckled honeycomb geometry have the novel physical properties akin to graphene such as the linear energy dispersion at the Fermi level. Besides, Silicene and Ge with the low buckled geometry and great SOC can be not only synthesized and processed using mature semiconductor techniques but also more easily integrated into the current electronics. All of these make silicene and Ge with the low buckled honeycomb geometry cornucopias of fundamental physics interests and promising applications.

\bibliography{Silicene}

{\bf Acknowledgements}

This work was supported by NSF of China (Grants No. 10974231) , the MOST Project of China (Grants No.2007CB925000, and 2011CBA00100) and Supercomputing Center of Chinese Academy of Sciences.

{\bf Author contributions}
Y.G.Y. conceived the idea and supervised the overall project.  Y.G.Y. and C.C.L. prepared the manuscript. C.C.L. carried out the main part of the calculation with assistance from Y.G.Y. and W.X.F.


{\bf Additional information}
 Correspondence and requests for materials should be addressed to Y.G.Y. (ygyao@aphy.iphy.ac.cn).

\end{document}